# First large-scale genomic prediction in the honey bee

Richard Bernstein[1], Manuel Du[1], Zhipei G. Du[1], Anja S. Strauss[1], Andreas Hoppe[1], and Kaspar Bienefeld[1]

[1] Institute for Bee Research Hohen Neuendorf, Friedrich-Engels-Str. 32, 16540 Hohen Neuendorf, Germany

Corresponding author: Richard Bernstein, Institute for Bee Research Hohen Neuendorf, Friedrich-Engels-Str. 32, 16540 Hohen Neuendorf, Germany. richard.bernstein@hu-berlin.de

# ABSTRACT

Genomic selection has increased genetic gain in several livestock species, but due to the complicated genetics and reproduction biology not yet in honey bees. Recently, 2 970 queens were genotyped to gather a reference population. For the application of genomic selection in honey bees, this study analyses the predictive ability and bias of pedigree-based and genomic breeding values for honey yield, three workability traits and two traits for resistance against the parasite *Varroa destructor*. For breeding value estimation, we use a honey bee-specific model with maternal and direct effects, to account for the contributions of the workers and the queen of a colony to the phenotypes. We conducted a validation for the last generation and a five-fold cross-validation. In the validation for the last generation, the predictive ability of pedigree-based estimated breeding values was 0.06 for honey yield, and ranged from 0.2 to 0.41 for the workability traits. The inclusion of genomic marker data improved these predictive abilities to 0.11 for honey yield, and a range from 0.22 to 0.44 for the workability traits. The inclusion of genomic data did not improve the predictive ability for the disease related traits. Traits with high heritability for maternal effects compared to the heritability for direct effects showed the most promising results. Across all traits, the bias with genomic methods was close to the bias with pedigree-based BLUP. The results show that genomic selection can successfully be applied to honey bees.

# 1. INTRODUCTION

Genomic selection (Meuwissen et al. 2001) incorporates genome-wide marker data into breeding value estimation. Compared to pedigree-based breeding values, the use of genomic data can increase the predictive ability of estimated breeding values, or enable the selection of animals before they are phenotyped. Both strategies have been realised to increase the genetic gain in several livestock species (Doublet et al. 2019; Fulton 2012; Samorè and Fontanesi 2016). Honey bee breeders, by contrast, employ phenotypic selection (De la Mora et al. 2020; Maucourt et al. 2020) or pedigree-based breeding value estimation (Bienefeld et al. 2007; Brascamp et al. 2016; Hoppe et al. 2020). Recently, a high density SNP chip was developed and genotypes of phenotyped queens are now available to validate genomic prediction (Jones et al. 2020).

Pedigree-based best linear unbiased prediction (**PBLUP**) of breeding values began in 1994 for the population registered on Beebreed. The estimated breeding values enabled hundreds of mostly Central European bee breeders to improve the quality of their stock (Hoppe et al. 2020). To ensure the quality of the estimated breeding values, the program relies on a specialized infrastructure for mating control and an adapted genetic model to account for the peculiarities of the honey bee (Bienefeld et al. 2007; Brascamp and Bijma 2014).

While the up to 50,000 workers in a hive do not reproduce under normal circumstances, they perform all other tasks in the hive, such as foraging, brood care, and cleaning (Koeniger et al. 2015). A queen mates with six to twenty drones during flight (Tarpy and Nielsen 2002). Therefore, the daughters of the queen will belong to various patrilines, and uncertain paternity is a challenge for the breeding value estimation. However, lasting breeding success in honey bees requires adequate measures for mating control (Plate et al. 2019). Therefore, mating stations are maintained with several drone producing queens (**DPQ**), which are usually unselected daughters of s single dam of high genetic quality.

The phenotypes of honey bee colonies for economically relevant traits result from the collaboration of worker groups and queens. In honey yield for example, the workers of a colony perform foraging and storing, but the queen affects the number of workers via her egg-laying rate, and influences the behavior of the workers via pheromones. Therefore, the genetic model for the traits includes direct and maternal effects for the contribution of workers and queens, respectively.

In commercial honey bee breeding programs, the demands of beekeepers lead to selection traits which differ significantly in terms of methodology and effort for recording and mathematical modelling. Typical aims include increased honey yield, better workability for the beekeeper, and more disease resistance (Petersen et al. 2020; Uzunov et al. 2017). Especially resistance against *Varroa destructor* is targeted, since this parasitic mite contributes to severe colony losses in numerous countries (Generschet al. 2010; Guichard et al. 2020; Traynor et al. 2016). Genomic breeding value estimation in honey bees has been tried in simulation studies, and single step genomic BLUP (**ssGBLUP**) appeared as an efficient solution (Bernstein et al. 2021; Gupta et al. 2013) to combine pedigree information with genomic information. The simulations showed that ssGBLUP can increase the accuracy of genomic breeding values considerably and enables high genetic gains, if the infrastructure is appropriately adapted. Augmenting ssGBLUP with trait-specific weights leads to weighted ssGBLUP (**WssGBLUP**) (Wang et al. 2012), which can increase the prediction accuracy further, as results from other species have shown (Lourenco et al. 2014; Teissier et al. 2019; Vallejo et al. 2019).

To our knowledge, only simulated results on genomic estimated breeding values in honey bees have been published until now. In this study, we firstly report the predictive abilities and the bias of PBLUP, ssGBLUP, and WssGBLUP for a number of key traits of economic importance in a large breeding population of honey bees.

# 2. MATERIALS AND METHODS

## Data

Pedigree and performance data from the *Apis mellifera carnica* population were used, since the genotyped queens belonged to this subspecies, which is native and widespread in Central Europe (Lodesani and Costa 2003; Ruttner 1988; Wallberg et al. 2014). The data was downloaded from BeeBreed on the 14th of February 2021, totalling 201 304 valid performance tests and pedigree data of 234 519 queens. The dataset was reduced and refined to the queens and performance tests relevant to compare classical and genomic selection, as follows. Performance tested queens on apiaries from test year 2010 on formed the set of potential phenotypes. Queens with a valid phenotype whose genotypes passed the quality control (see below) defined the seed. In an iterative process, phenotypes were added by (1) the completion of testing apiaries, (2) the completion of sister groups, and (3) the closing of pedigree gaps, until no further phenotypes could be added. Finally, the full ancestry of all resulting queens was added without phenotype. The final enriched dataset contained 36 509 phenotypes in a pedigree of 44 183 queens and 4 512 sires composed of DPQ. Table 1 lists the countries of origin for all colonies.

**Table 1** Number of phenotyped and genotyped queens included in the data set by country.

| Country | Phenotyped queens | Genotyped queens after quality control |
|---|---|---|
| Germany | 24 019 | 1 982 |
| Austria | 9 618 | 372 |
| Italy | 796 | 1 |
| Switzerland | 619 | 17 |
| Ukraine | 467 | 0 |
| Belgium | 368 | 4 |
| Netherlands | 275 | 11 |
| Sweden | 133 | 0 |
| France | 117 | 0 |
| Croatia | 91 | 2 |

The phenotypes covered honey yield, gentleness, calmness, swarming drive, hygienic behavior, and *Varroa* infestation development (**VID**). Honey yield was measured in kg, and the values were corrected for outliers as described in (Hoppe et al. 2020). Gentleness, calmness, and swarming tendency were recorded as marks from 1 to 4 where 4 is best. Records for these traits were discarded if all colonies on an apiary received the same mark. For hygienic behavior, larvae were artificially killed with a pin and the percentage of cleared cells was recorded (Büchler et al. 2013). VID indicates the resistance of a colony against *Varroa*, based on the change of the level of *Varroa* infestation from early spring to late summer (see (Hoppe et al. 2020) for the calculation of VID). For a measurement of *Varroa* infestation, a bee sample is taken from the hive, and the number of mites per 10g bees is determined (Büchler et al. 2013). Table 2 shows the descriptive statistics of the phenotypes available for each trait.

**Table 2** Descriptive statistics for honey yield, gentleness, calmness, swarming drive, hygienic behavior, and *Varroa* infestation development (VID).

| Trait | Number of records | Number of genotyped queens with record | Average size of apiaries with a genotyped queen (SD) | Mean | SD | Min. | Max. |
|---|---|---|---|---|---|---|---|
| Honey yield | 35 888 | 2 046 | 13.62 (8.33) | 40.71 | 22.84 | 0 | 199.8 |
| Gentleness | 35 187 | 2 013 | 13.80 (8.48) | 3.52 | 0.48 | 1 | 4 |
| Calmness | 34 652 | 2 016 | 13.76 (8.50) | 3.49 | 0.48 | 1 | 4 |
| Swarming drive | 26 937 | 1 549 | 14.57 (8.88) | 3.55 | 0.76 | 1 | 4 |
| Hygienic behavior | 23 924 | 1 781 | 13.36 (7.86) | 62.26 | 23.13 | 0 | 100 |
| VID | 24 650 | 1 787 | 13.48 (7.85) | -1.55 | 2.38 | -77.12 | 6.93 |

**Legend.** Honey yield is given in kg. Marks from 1 to 4 were recorded for gentleness, calmness, and swarming drive. Hygiene is given as the percentage of cleared cells. VID is a *Varroa* resistance score and higher values indicate more resistance.

The 100-K-SNP chip (Jones et al. 2020) was used to genotype 2 970 queens which were registered on BeeBreed and born between 2009 and 2017. Markers which were called in less than 90% of the samples, had minor allele frequency below 1%, or showed significant deviations from Hardy-Weinberg-equilibrium after Bonferroni-correction (chi-square p-value $< 0.05 \times 10^{-5}$) were removed. This left 63 240 markers for further analysis. 312 queens were removed because less than 90% of all the valid markers were called in their samples. After comparisons of daughter and parent based on the number of opposing homozygotes, 207 queens were removed. Subsequently, 62 samples were removed based on comparison of genomic and classic relationship matrix (Calus et al. 2011). This left 2 389 genotyped queens for further analysis.

## Model and genetic parameters

The complex collaboration between the workers and the queen of a colony must be reflected in the model, and carefully analyzed in the calculation of genetic parameters (Brascamp and Bijma 2019). The phenotype, $y$, of a colony is modeled as

$$y = a_W + m_Q + e, \qquad (1)$$

where $a_W$ is the direct effect of the worker group in the colony, and $m_Q$ the maternal effect of the queen in the colony, while $e$ is a non-heritable residual.

The phenotypic variance was calculated according to formula (2) in (Brascamp and Bijma 2019) as

$$\sigma_{ph}^2 = A_{base}\sigma_a^2 + \sigma_m^2 + \sigma_{am} + \sigma_e^2, \qquad (2)$$

where $\sigma_a^2$ and $\sigma_m^2$ are the additive genetic variances of direct and maternal effects, $\sigma_{am}$ is the covariance between direct and maternal effects, $\sigma_e^2$ is the residual variance, and $A_{base}$ is the average relationship between two workers of the same colony in the base population. The variance components were estimated via AIREML with the complete phenotypic information, using the model for PBLUP (see below). We used $A_{base} = 0.40$ (Brascamp and Bijma 2019), because even the oldest queens in our pedigree came from populations with established mating control (Armbruster 1919). The heritabilities of direct and maternal effects, and the total heritability, $h_a^2$, $h_m^2$, and $h_T^2$ were calculated according to formulas (6b), (6c), and (7c) in (Brascamp and Bijma 2019), respectively, as

$$h_a^2 = A_{base}\sigma_a^2/\sigma_{ph}^2, \quad h_m^2 = \sigma_m^2/\sigma_{ph}^2 \text{ and } h_T^2 = \frac{\sigma_a^2 + \sigma_m^2 + 2\sigma_{am}}{\sigma_{ph}^2}. \qquad (3)$$

However, because the Beebreed dataset relies on colony based selection (CBS), we calculate the heritability of the corresponding selection criterion, $h^2_{CBS}$ (derived as formula (6) in (Bernstein et al. 2021); called accessible heritability in (Hoppe et al. 2020)), as

$$h^2_{CBS} = A_{base} h^2_T. \qquad (4)$$

# Breeding value estimation

The following mixed linear model was used for PBLUP.

$$\mathbf{y} = \mathbf{Xb} + \mathbf{Z}_a \mathbf{a} + \mathbf{Z}_m \mathbf{m} + \mathbf{e}, \qquad (5)$$

where $\mathbf{y}$ is a vector of observations on colonies; $\mathbf{b}$ a vector of fixed effects (year and apiary); $\mathbf{a}$ a vector of direct effects of queens, worker groups or sires; $\mathbf{m}$ a vector of maternal effects of queens, worker groups or sires; $\mathbf{e}$ a vector of residuals; and $\mathbf{X}$, $\mathbf{Z}_a$, and $\mathbf{Z}_m$ are known incidence matrices for $\mathbf{b}$, $\mathbf{a}$, and $\mathbf{m}$, respectively. The expected values of $\mathbf{a}$, $\mathbf{m}$, and $\mathbf{e}$ were assumed to equal $\mathbf{0}$, with the following variances:

$$\mathrm{Var} \begin{pmatrix} \mathbf{a} \\ \mathbf{m} \\ \mathbf{e} \end{pmatrix} = \begin{pmatrix} \sigma_a^2 \mathbf{A} & \sigma_{am} \mathbf{A} & 0 \\ \sigma_{am} \mathbf{A} & \sigma_m^2 \mathbf{A} & 0 \\ 0 & 0 & \sigma_e^2 \mathbf{I} \end{pmatrix}, \qquad (6)$$

where $\mathbf{A}$ is the honey bee specific numerator relationship matrix derived from pedigree (Brascamp and Bijma 2014), $\mathbf{I}$ is an identity matrix, and $\sigma_a^2$, $\sigma_m^2$, $\sigma_{am}$ and $\sigma_e^2$ are the additive genetic variance of worker and queen effects, their covariance and the residual variance, respectively.

The model equation and variances for ssGBLUP were the same as for PBLUP, except for the fact that matrix $\mathbf{H}$ replaced matrix $\mathbf{A}$. Matrix $\mathbf{H}$ was constructed from the numerator relationship matrix $\mathbf{A}$ which is calculated from pedigree information, and the marker information in the following steps (Aguilar et al. 2010; Christensen and Lund 2010). The genomic relationship matrix , $\mathbf{G}$, ((VanRaden 2008), method 1) was constructed by the following equation.

$$\mathbf{G} = \frac{\mathbf{ZZ}^T}{2\sum_i p_i(1-p_i)}, \quad (7)$$

where $p_i$ is the allele frequency of the SNP at locus $i$; $\mathbf{Z} = \mathbf{M} - \mathbf{P}$ with $\mathbf{M}$ containing the marker information of all genotyped queens given as 0, 1, 2, and matrix $\mathbf{P}$ defined column-wise by $P_{ij} = 2p_i$ for all $j$. Matrix $\mathbf{G}$ was adjusted to $\mathbf{A}$ by adjusting the means of diagonal and off-diagonal elements as described by (Christensen et al. 2012). To have an invertible genomic relationship matrix, we used the weighted genomic relationship matrix, $\mathbf{G}_w$, given by the following equation.

$$\mathbf{G}_w = 0.95\mathbf{G} + 0.05\mathbf{A}_g, \quad (8)$$

where $\mathbf{A}_g$ is the submatrix of $\mathbf{A}$ relating to the genotyped animals. Finally, the inverse of $\mathbf{H}$ was computed according to the following formula.

$$\mathbf{H}^{-1} = \mathbf{A}^{-1} + \begin{pmatrix} 0 & 0 \\ 0 & \mathbf{G}_w^{-1} - \mathbf{A}_g^{-1} \end{pmatrix}. \quad (9)$$

The model equation and variances for WssGBLUP were the same as for ssGBLUP, except for the fact that matrix $\mathbf{G}^*$ replaced matrix $\mathbf{G}$. Matrix $\mathbf{G}^*$ was constructed from the vectors of direct and maternal additive genetic effects, $\mathbf{a}$ and $\mathbf{m}$, and the genomic relationship matrix $\mathbf{G}_w$, which were obtained from ssGBLUP. The vectors of the direct and maternal SNP effects, $\mathbf{u}$ and $\mathbf{v}$, were estimated by:

$$\mathbf{u} = \lambda \mathbf{M}^T \mathbf{G}_w^{-1} \mathbf{a}, \quad (10)$$

$$\mathbf{v} = \lambda \mathbf{M}^T \mathbf{G}_w^{-1} \mathbf{m},$$

with $\lambda = \frac{1}{2\sum_i p_i(1-p_i)}$, where $p_i$ and $\mathbf{M}$ have the same value as in ssGBLUP. SNP weights $\mathbf{d}$ were calculated using the average of the direct and maternal SNP effects, deviating from the original algorithm which considered only single-trait models (Wang et al. 2012) as follows.

$$d_i = \left(\frac{u_i + v_i}{2}\right)^2 2p_i(1-p_i). \quad (11)$$

Diagonal matrix **D** was defined by $D_{ii} = d_i/\overline{\mathbf{d}}$, where $\overline{\mathbf{d}}$ is the average of **d**. The trait-specific matrix $\mathbf{G}^*$ was calculated by the following formula.

$$\mathbf{G}^* = \frac{\mathbf{ZDZ}^T}{2\sum_i p_i(1-p_i)}, \qquad (12)$$

where **Z** is the same matrix as in ssGBLUP.

Programs from the BLUPf90 software (Misztal et al. 2002) were used to estimate the genetic parameters, predict breeding values and calculate relationship matrices **G** and $\mathbf{G}^*$. To account for the specifics of honey bees, PInCo (Bernstein et al. 2018) was used to calculate the pedigree-based relationship matrices. Equations (8) to (11) were implemented in R (R Development Core Team 2020).

## Validation

We performed two types of cross-validation. In the generation validation, estimated breeding values (**EBV**) were predicted using PBLUP, ssGBLUP and WssGBLUP (1) without the phenotypes of all queens born in 2017 or later, and (2) without the phenotypes of queens born in 2016 or later. The EBV of the 265 genotyped queens born in 2017 from scenario 1 were pooled with EBV of the 994 genotyped queens born in 2016 from scenario 2.

In the five-fold cross-validation, only apiaries with at least five performance tested queens were included to ensure reliable estimates of fixed effects. This left 1 281 genotyped queens for the validation. The 1 281 queens were split into five partitions, where apiaries were evenly distributed onto the partitions. For each partition, EBV were estimated using PBLUP, ssGBLUP and WssGBLUP without the phenotypes of the animals on this partition, and the results from all partitions were pooled. The procedure was repeated six times.

To assess the predictive ability of PBLUP, ssGBLUP and WssGBLUP, *predicted phenotypes* were correlated to *phenotypes corrected for fixed effects*, where the predicted phenotype of a colony is the sum of the direct effect of its worker group and the maternal effect of its queen. For each method to predict EBV, the phenotypes corrected for fixed effects were calculated using fixed effects from the same method. In the generation-validation, PBLUP, ssGBLUP and WssGBLUP were run on the complete data set to obtain appropriate fixed effects. In the five-fold cross-validation, the fixed effects for the correction of the phenotypes were taken from the same run of the same partition as the predicted phenotypes.

A bootstrap procedure was used to test whether the predictive abilities of WssGBLUP and ssGLUP were significantly higher than the predictive ability of PBLUP. 10,000 bootstrap sample vectors were constructed by sampling validation queens with replacement, and the predictive ability with PBLUP, ssGLUP, and WssGBLUP was calculated for each vector. Two methods were considered significantly different, if the same method had higher predictive ability in 97.5% of all sample vectors (p-value of 0.05 in a two-sided test). Similar bootstrapping methods were used in other studies (Iversen et al. 2019; Legarra et al. 2008).

The regression coefficient of *realized total EBV of queens* on *predicted total EBV of queens*, $b_1$, was used as a measure of bias. Values of $b_1 < 1$ and $b_1 > 1$ indicate inflation and deflation of the predicted breeding values compared to the realized total EBV, respectively. For both the generation validation and the five-fold cross-validation, the realized total EBV of a queen and its predicted total EBV were the sum of its direct and maternal effect calculated using the complete data set, and restricted data, respectively. The predicted total EBV of PBLUP, ssGBLUP, and WssGBLUP were compared to realised total EBV from the same method.

# 3. RESULTS

## Genetic parameters

Estimates of the genetic parameters are shown in Table 3. The total heritability was very high for gentleness and calmness, medium for hygienic behavior, honey yield and swarming drive, low for VID. All traits showed considerable negative genetic correlations between maternal and direct effects. The heritability for direct effects was considerably larger than the heritability for maternal effects in gentleness, calmness, and hygienic behavior, but equal to or smaller than the heritability for maternal effects for all other traits.

**Table 3.** Estimated variance and covariance components, genetic parameters derived from these (co)variances. The approximate standard deviations are given in brackets.

| Trait | $\sigma_a^2$ | $\sigma_m^2$ | $\sigma_{am}$ | $\sigma_e^2$ | $h_a^2$ | $h_m^2$ | $r_G$ | $h_T^2$ | $h_{CBS}^2$ |
|---|---|---|---|---|---|---|---|---|---|
| Honey yield | 27.235 (5.406) | 13.841 (3.015) | -6.124 (3.372) | 61.549 (1.476) | 0.136 (0.027) | 0.173 (0.037) | -0.315 (0.142) | 0.360 (0.047) | 0.144 (0.019) |
| Gentleness | 0.134 (0.016) | 0.027 (0.006) | -0.020 (0.008) | 0.062 (0.003) | 0.435 (0.053) | 0.221 (0.049) | -0.325 (0.096) | 0.991 (0.083) | 0.396 (0.033) |
| Calmness | 0.103 (0.013) | 0.019 (0.005) | -0.013 (0.006) | 0.059 (0.003) | 0.387 (0.048) | 0.181 (0.043) | -0.289 (0.105) | 0.907 (0.076) | 0.363 (0.03) |
| Swarming drive | 0.143 (0.033) | 0.054 (0.017) | -0.008 (0.019) | 0.356 (0.010) | 0.124 (0.029) | 0.117 (0.037) | -0.087 (0.249) | 0.394 (0.056) | 0.158 (0.022) |
| Hygienic behavior | 111.61 (22.027) | 25.571 (8.508) | -5.454 (10.688) | 174.73 (5.688) | 0.186 (0.037) | 0.107 (0.035) | -0.102 (1.645) | 0.527 (0.064) | 0.211 (0.026) |
| VID | 0.159 (0.045) | 0.068 (0.024) | -0.028 (0.027) | 0.619 (0.014) | 0.088 (0.025) | 0.095 (0.033) | -0.270 (0.275) | 0.236 (0.044) | 0.095 (0.018) |

**Legend.** VID, *Varroa* infestation development; the last nine columns show the additive genetic variances of direct ($\sigma_a^2$) and maternal effects ($\sigma_m^2$), their covariance ($\sigma_{am}$), the residual variance ($\sigma_e^2$), the heritabilities of direct effects ($h_a^2$), maternal effects ($h_m^2$), the genetic correlation ($r_G$), the total heritability ($h_T^2$), and the heritability for the selection criterion in colony based selection ($h_{CBS}^2$).

# Accuracy of breeding values

The predictive abilities of the methods under investigation in the generation validation are shown in Figure 1. Compared to PBLUP, the predictive ability was improved with WssGBLUP for honey yield (94%), swarming drive (7%), gentleness (6%), calmness (5%), and VID (20%), and with ssGBLUP, improvements were observed for honey yield (48%), VID (41%), and gentleness (6%). The improvement with WssGBLUP over PBLUP for honey yield was statistically significant. No improvement was observed for hygienic behavior, and ssGBLUP did not yield a higher accuracy than PBLUP for calmness and swarming drive.

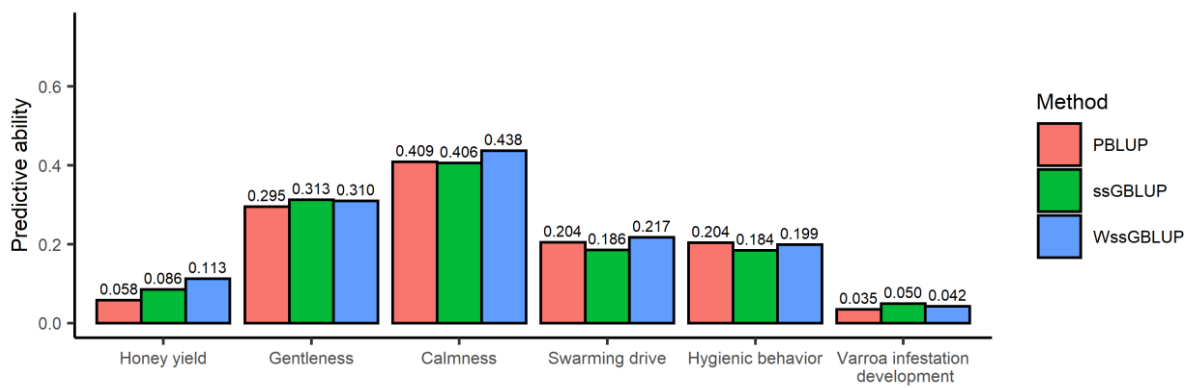

**Figure 1.** Predictive abilities of pedigree-based BLUP (PBLUP), single step genomic BLUP (ssGBLUP) and weighted ssGBLUP (WssGBLUP) in the generation validation.

The accuracies of the methods under investigation in the five-fold cross-validation are shown in Figure 2. Improvements over PBLUP were achieved for swarming drive (20%), honey yield (15%), calmness (2%), and gentleness (3%) with WssGBLUP. Improvement over PBLUP with ssGBLUP were achieved for honey yield (10%), and swarming drive (3%). The improvements with WssGBLUP over PBLUP were statistically significant for calmness and swarming drive. No improvement was observed for hygienic behavior and VID.

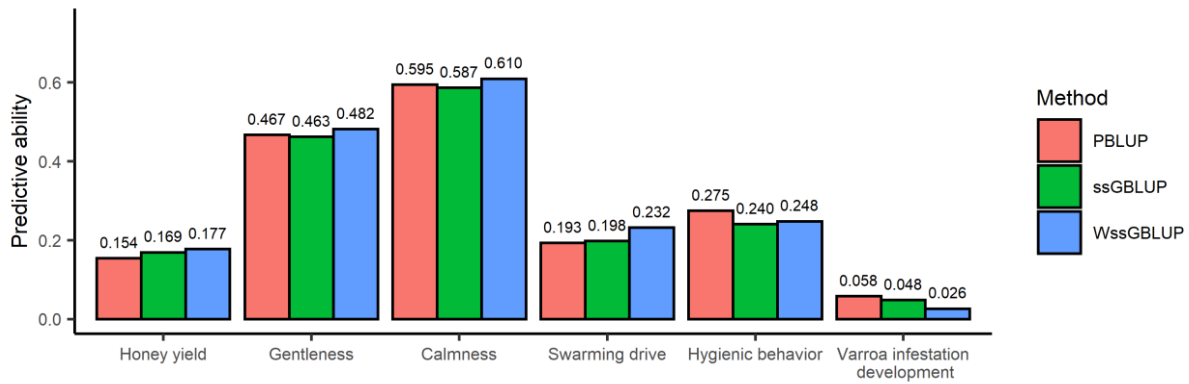

**Figure 2.** Mean Predictive abilities of pedigree-based BLUP (PBLUP), single step genomic BLUP (ssGBLUP) and weighted ssGBLUP (WssGBLUP) in the five-fold cross-validation, calculated across the six repetitions. The standard deviations over the six repetitions are not shown, as they were smaller than 0.007.

Overall, both validations showed similar results, although the predictive ability was higher in the five-fold cross-validation, and the increases in predictive ability with ssGBLUP and WssGBLUP over PBLUP were higher in the generation validation.

## Bias of breeding values

Bias was calculated as the regression coefficient $b_1$ of total realized EBV on total predicted EBV. The results for EBV from PBLUP, ssGBLUP and WssGBLUP in the generation validation are shown in Figure 3. The regression coefficient $b_1$ deviated the most from 1 for honey yield and VID with -0.33, and -0.28, respectively, for PBLUP. The results for WssGBLUP, and ssGBLUP are rather similar to each other, and more stable across the traits than with PBLUP. With WssGBLUP, the regression coefficient $b_1$ deviated from 1 by a number between -0.13 and -0.26 across all traits. The results for all three methods show inflated EBV estimates.

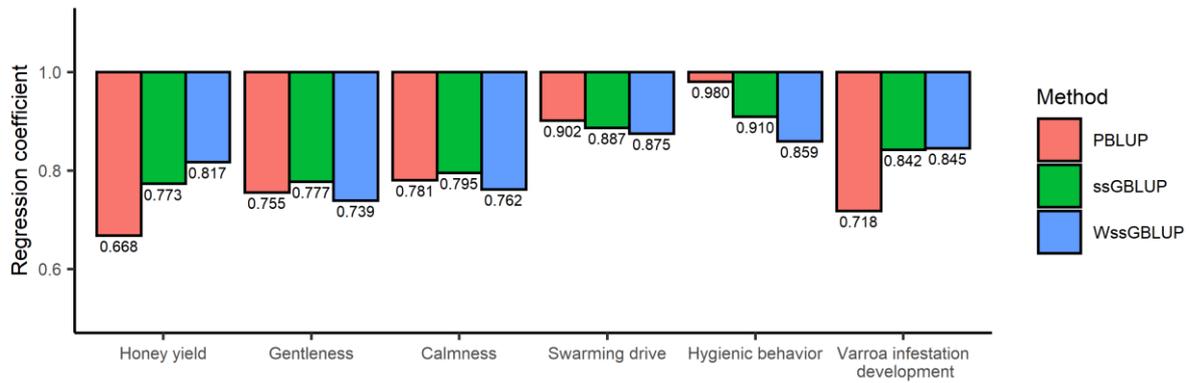

**Figure 3.** Regression coefficient $b_1$ of pedigree-based BLUP (PBLUP), single step genomic BLUP (ssGBLUP) and weighted ssGBLUP (WssGBLUP) in the generation validation.

The results for EBV from PBLUP, ssGBLUP and WssGBLUP in the five-fold cross-validation are shown in Figure 4. For PBLUP, the regression coefficient $b_1$ deviated from 1 by a number between -0.20 and -0.36. The deviations of $b_1$ from 1 for ssGBLUP were slightly higher, and WssGBLUP showed the highest bias for all traits. All three methods showed inflated EBV estimates.

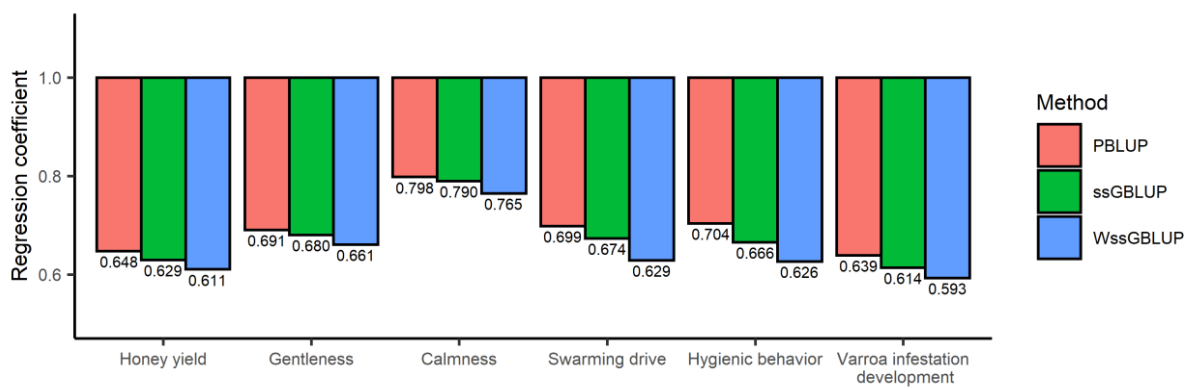

**Figure 4.** Mean Regression coefficient $b_1$ of pedigree-based BLUP (PBLUP), single step genomic BLUP (ssGBLUP) and weighted ssGBLUP (WssGBLUP) in the five-fold cross-validation, calculated across the six repetitions. The standard deviations over the six repetitions are not shown, as they were smaller than 0.004.

# 4. DISCUSSION

## Genetic Parameters and Predictive Ability

The estimated genetic parameters (Table 3) are in line with the results for the multiple trait models of the complete Beebreed data set (Hoppe et al. 2020). The results on the predictive abilities in the generation validation (Figure 1) and in the five-fold cross-validation (Figure 2) show improvements with WssGBLUP over PBLUP for honey yield, gentleness, calmness and swarming drive. These results were within the range reported for data sets of similar size in dairy goats (Legarra et al. 2014), or for traits affected by maternal effects in beef cattle (Lourenco et al. 2015) or pigs (Putz et al. 2018).

The results on the difference in predictive ability between WssGBLUP and PBLUP can be explained with the results on the heritabilities (Table 3). Because simulation studies in honey bees showed greater increases in predictive ability with ssGBLUP over PBLUP for maternal effects than for direct effects (Bernstein et al. 2021), traits with a higher heritability for maternal effects than for direct effects can be expected to show higher increases than other traits in predictive ability with WssGBLUP and ssGBLUP over PBLUP. Honey yield and swarming drive showed the highest improvements in predictive ability with WssGBLUP over PBLUP, and the heritability for maternal effects is equal to or greater than the heritability for direct effects in both traits. Although the heritability for maternal effects is also equal to the heritability for direct effects in VID, the results for this trait are rather ambiguous. This is due to the low total heritability for this trait, because simulation studies in honey bees and other species show that traits with low heritability also have low accuracy of pedigree-based and genomic EBV (Gowane et al. 2019; Gupta et al. 2013). This result stands out from other species where maternal effects are modelled, as in beef cattle (Lourenco et al. 2018) and simulation studies for beef cattle and pigs (Lourenco et al. 2013; Putz et al. 2018), the predictive ability for direct effects showed higher increases in predictive ability with ssGBLUP over PBLUP than the predictive ability for maternal effects.

An analogy between direct effects in honey bees and traits in other species can explain, why traits with a higher heritability for direct effects than for maternal effects (e.g. hygienic behavior) showed less increase in predictive ability from PBLUP to WssGBLUP than other traits of similar total heritability (e.g. swarming drive). For our study, only queens were genotyped, since they are the candidates of selection. The primary selection candidates in dairy cattle or pigs are males. For traits for which only females are phenotyped, simulations showed that male selection candidates should be genotyped to achieve high genetic gain, but including additional genotypes of females increases the accuracy of prediction considerably (Buch et al. 2012; Lillehammer et al. 2011; Plieschke et al. 2016). The situation of the direct effects of queens and workers in our study is similar to the situation of males and females in dairy cattle and specific traits in pigs, respectively, since direct effects quantify the impact of workers on the phenotypes of colonies. The results on dairy cattle and pigs suggest that genotyping queens has the highest priority, but including genotypes of workers might increase the accuracy of genomic prediction, especially for direct effects. To our knowledge, there are no studies in honey bees on this subject. However, since the workers of a hive belong to different patrilines, the price for genotyping enough workers to represent the workers as a whole is probably too high in practice.

The results for the *Varroa* resistance related traits were also by problems in gathering data. The number of genotyped queens with phenotype for both traits was about 200 queens lower than for honey yield, gentleness, and calmness. Furthermore, the number of phenotyped queens on apiaries with a genotyped queen (Table 2) was low for the *Varroa* related traits, which might have led to less accurate fixed effects. However, *Varroa* specific hygienic behavior is the subject of ongoing research (Conlon et al. 2019; Farajzadeh et al. ; Mondet et al. 2020). The discovery of new quantitative trait loci (QTL) which are then covered by causative SNPs on a new chip can increase predictive ability for the *Varroa* related traits considerably.

The predictive ability of ssGBLUP was slightly lower than the predictive ability of WssGBLUP for most traits. This result is common in studies for several other agricultural species using WssGBLUP (e.g. (Lu et al. 2020; Teissier et al. 2019; Wang et al. 2014)). In simulation studies (Lourenco et al. 2017; Wang et al. 2012), WssGBLUP had higher predictive ability than ssGBLUP when the trait was controlled by few QTL, and both methods showed equal predictive ability when the trait was polygenic. As the predictive ability for VID was higher with ssGBLUP than with WssGBLUP in both validations, the genetic architecture of the trait appears to be highly polygenic. However, this is a preliminary conclusion, as VID has the lowest heritability of the traits we considered, due to the many factors that affect it (see (Guichard et al. 2020) for a review).

The predictive abilities in the five-fold cross-validation were for the majority of the traits higher than in the generation validation. This is due to the fact that in the five-fold cross-validation, sibling groups are evenly distributed across the partitions, while the phenotypes of whole sibling groups might be removed for the calculation of EBVs in the generation-validation. Therefore, the five-fold cross-validation is a validation within sibling groups, while the generation validation is similar to a validation across sibling groups. Studies in other species found that validations within sibling groups show higher predictive abilities than validations across siblings groups (Gao et al. 2019; Kjetså et al. 2020; Legarra et al. 2008). The standard deviations of the predictive abilities in the five-fold cross-validation were extremely small in our study, but the predictive abilities for individual partitions showed large differences.

According to a simulation study in honey bees (Bernstein et al. 2021), the size of the reference population in our study is close to the minimal size which should be available to initiate a breeding program. We expect the reference population to grow in the future, when breeders start to apply genomic selection.

The larger reference population is likely to obviate the need to run WssGBLUP instead of ssGBLUP, since a simulation study showed that WssGBLUP and ssGBLUP yield the same results for large reference sets (Lourenco et al. 2017). The larger reference population will also result in an increase of the predictive ability of genomic methods, as results from other species demonstrate (Daetwyler et al. 2012; Lourenco et al. 2015; Mehrban et al. 2017; Moser et al. 2009).

## Bias of the estimated breeding values

The regression coefficients, $b_1$ of realized total EBV of queens on predicted total breeding values of queens are shown in Figures 3 for the generation validation. The results for WssGBLUP deviate slightly further than -0.15 from 1 for honey yield, gentleness, and calmness. As 0.15 is seen as the maximum deviation from 1 for acceptable EBV (Tsuruta et al. 2011), our results for these traits are biased. The results of the five-fold cross-validation (Figure 4) confirm that the estimates show high inflation. The difference between the validations is due to the different ways of accounting for fixed effects. In the generation validation, the same value of the fixed effect was modelled for all colonies on the same apiary in a single year, while in the five-fold cross-validation different values of the fixed effect for colonies on the same apiary in a single year were used, depending on the partition. This difference was exacerbated by the small apiaries in honey bees, which lead to a higher variance of the predicted EBV in the five-fold cross-validation than in the generation validation, and consequently lower regression coefficients. Results from routine evaluations are expected to be closer to the generation validation, where the bias was acceptable for swarming drive, hygienic behavior, and VID, and close to acceptable for the remaining traits.

The bias with genomic methods can be reduced by e.g. increasing the share of the classic relationship matrix $\mathbf{A}_g$ in equation (8) (McMillan and Swan 2017; Misztal et al. 2017).

However, considerable bias was neither observed in simulations for honey bees (Bernstein et al. 2021) for PBLUB and ssGBLUP, nor the Austrian data set (Brascamp et al. 2016) with PBLUP. Since, our data set is a small outtake of a very large population, we expect that PBLUP and genomic breeding values will show less bias as the data set increases.

## Practical application of genomic selection in the honey bee

Gathering genomic data from honey bees for genomic selection requires special considerations, due to their small body size, and their genetic diversity within a hive. Non-lethal ways to genotype queens are available for genomic selection (Jones et al. 2020). The exuviae which queens leave behind after hatching offer a non-lethal option to genotype virgin queens, but just one exuvia is available for each queen, and exuviae showed low DNA quality in several cases. Alternatively, drones can be gathered from a hive to genotype the queen, since drones are haploid offspring. This method is a viable option to genotype queens before the colony is phenotyped.

The availability of genomic breeding values offers new possibilities in breeding schemes for honey bees. Queens can be genomically preselected by genotyping candidate queens, and keeping only the ones with the highest genomic breeding value for phenotyping or deployment as DPQ on mating stations. A simulation study suggests that genomic preselection can increase the genetic gain per year considerably, if at least 1 000 queens per year are genotyped and later phenotyped (Bernstein et al. 2021). Another simulation study suggests, that if the structure of the breeding program is changed so that several generations of queens are genomically preselected in a single summer, the generation interval will at least be halved, enabling even higher genetic gains (Brascamp et al. 2018).

The routine evaluations of queens on a SNP chip can also benefit beekeepers by checking the subspecies of the queens and monitoring the genetic diversity within the population.

Conservation efforts have been initiated for several subspecies in Europe (De la Rúa et al. 2009; Janczyk et al. 2021; Pinto et al. 2014), at least in part because locally adapted ecotypes have a higher survivability (Büchler et al. 2014). While microsatellites offer a cheap and widely used way to monitor genetic diversity in honey bees (Meixner et al. 2013), a sufficiently high number of SNPs is more accurate when the SNPs are properly validated (Muñoz et al. 2017).

# 5. CONCLUSIONS

WssGBLUP offers significantly greater predictive ability than PBLUP for honey yield, calmness, and swarming drive. For gentleness, the predictive ability of WssGBLUP was greater than the predictive ability of PBLUP to a similar degree as for calmness, but the difference remained below the threshold for significance. Across all traits, the bias with WssGBLUP and ssGBLUP was close to the bias with PBLUP. However, ssGBLUP offers too little improvement over PBLUP to be recommended based on the current data set for *Varroa* resistance traits, which is likely due to the size of the reference population. A larger reference population or the discovery of new causative SNPs for *Varroa* resistance are required to increase the predictive ability of genomic methods for hygienic behavior and VID. The results suggest that genomic selection can be successfully applied to honey bees.

## Acknowledgements

This work is part of the research program "Establishment of genomic selection in order to improve disease resistance, performance, behavior, and genetic diversity in the honeybee" with project number 742 397. The project is supported by funds of the German Government's Special Purpose Fund held at Landwirtschaftliche Rentenbank. Additional funding was provided by the European Commission under its FP7 KBBE program (2013.1.3-02, for project SmartBees Grant Agreement number 613960), the Deutsche Forschungsgemeinschaft (DFG,

German Research Foundation, Grant number 462225818), and the German federal states of Brandenburg, Berlin, Sachsen, Sachsen-Anhalt and Thüringen.

# Author contribution statement

RB, ZD, and AS prepared the genotype data. AH prepared phenotype and pedigree data. RB implemented the estimation of breeding values and genetic parameters, analysed the results, and wrote the manuscript. MD, ZD, AS, AH, KB assisted with the interpretation of the results, and writing the manuscript. AS, AH and KB supervised the study. All authors read and approved the manuscript.

# Conflict of interest

The authors declare that they have no competing interests.

# Data archiving

The genotypes used for this study are available in (Jones et al. 2020). The phenotype data of this study belongs to several breeding associations and is unavailable due to legal reasons. Requests to access further raw material should be directed at the authors of this study.